\documentclass[aps,prc,twocolumn,superscriptaddress,10pt]{revtex4-1}

\usepackage{graphicx}
\usepackage{dcolumn}
\usepackage[version=4]{mhchem}
\usepackage{amsmath}
\usepackage{amssymb}

\usepackage[english]{babel} 
\usepackage{url} 
\usepackage[colorlinks,
linkcolor=blue,
anchorcolor=blue,
urlcolor=blue,
citecolor=blue]{hyperref} 
\begin{document}

\title{Microscopic Description of Rotational Nuclear Fission Elucidates Fragment Spin Generation and Scission Mechanism}

\author{Yu Qiang}
\affiliation{Hangzhou International Innovation Institute of  Beihang University,  Hangzhou 311100, China}
\affiliation{School of Physics, and State Key Laboratory of Nuclear Physics and Technology, Peking University, Beijing 100871, China}

\author{Zhibo Li}
\affiliation{School of Physics, and State Key Laboratory of Nuclear Physics and Technology, Peking University, Beijing 100871, China}

\author{Junchen Pei}\email[]{peij@pku.edu.cn}
\affiliation{School of Physics, and State Key Laboratory of Nuclear Physics and Technology, Peking University, Beijing 100871, China}
\affiliation{Southern Center for Nuclear-Science Theory (SCNT), Institute of Modern Physics, Chinese Academy of Sciences, Huizhou 516000, China}

\begin{abstract}
The generation of fission fragment spin as a probe of scission mechanism remains a question of considerable interests.
We present here  microscopic calculations of rapidly rotating fission of the compound nucleus $^{240}$Pu under varying initial conditions within the time-dependent density functional theory framework.
The obtained spin
ratio of light to heavy nascent fragments is unchanged up to high spins but diminished as excitation energies increase,
indicating sawtooth structures in fragment spins would fade away at high excitations rather than high angular momentum.
Further analysis elucidates that the bending scission mode is predominated at low excitation energies, which has been under intense debates.
Results also show thicker and elongated neck configurations, along with scission nucleons emitted perpendicular to the fission axis,  owing to rapid rotations.
These findings offer insights into the scission mechanism of rotating compound nuclei that have usually been overlooked in earlier studies.
\end{abstract}

\maketitle





\emph{Introduction.}---
Nuclear scission in the final stages of fission is not a trivial splitting along
the fission axis,
as indicated by considerable fragment spins  about 7$\hbar$ generated even in
spontaneous fission~\cite{exp_frag_spin1}. This is distinctly different from the controllable separation between clouds of atoms~\cite{clouds}.
In 2021, an accurate measurement of fission fragment (FF) intrinsic spins
shows surprising sawtooth structures of fragment spin distributions as a function of fragment masses~\cite{Wilson}.
The similar sawtooth structures have been observed in distributions of neutron multiplicities, which
is related to the energy partition between binary fragments~\cite{NeutSaw1,NeutSaw2,carlsson}.
Furthermore, the experiment found no spin correlations between fragments and FF spins
are likely to be generated after scission. Subsequently, strong interests  have
been motivated, however, the generation of  FF spin and the scission mechanism
are still under debates~\cite{StetcuL,RandrupL,RandrupC1,RandrupC4,BulgacL1,BulgacL2,BulgacC1,ScampsC1,ScampsC2,photon_exp}.

The deeper understanding of nuclear scission is crucial for the application needs of high quality multiple post-fission data.
The intuitive modeling of nuclear scission as a random rupture in the neck has been widely adopted
in descriptions of FF yields~\cite{brosa,scission_point}. The microscopic time-dependent density functional theory (TDDFT) reveals that
the scission is a fast non-adiabatic process so that quantum entanglement between two separated fragments is persistent,
impacting the partition of particles and energies between fragments~\cite{qy4,shang}.
The FF spin can be generated by specific scission modes, mainly the
bending and wriggling modes in which FF spins are perpendicular to the fissioning axis~\cite{mode,mode2}.
TDDFT calculations claim the bending mode is predominated~\cite{BulgacL2}, in contrast to the wriggling
mode by FREYA as a Monte Carlo modeling~\cite{freya_mode}.

The FF spins inherited from initial angular momentum of compound nuclei (CN)
have usually not been taken into account in existing models, as they focus on fissioning nuclei with low angular momentum~\cite{BulgacL2,ScampsC1,Marevic1,RandrupC1}.
 FF spins with low initial excitation energies and low angular momentum
are determined by  spin fluctuations at scission~\cite{RandrupC2}.
 However, it is of interests to know the scission mechanism of rapidly
rotating nuclei, and consequently the partition
of angular momentum between fragments, with both inherited and fluctuation spins.
The fission dynamics is much related to excitation energies, leading
to enhanced viscosity~\cite{qy2} and the fade away of sawtooth structures~\cite{qy4,qy1,carlsson}.
Both TDDFT and FREYA results show that the dependence of FF spin on the angular momentum of CN is weak
and is more related to excitation energies~\cite{BulgacL1,RandrupC2}.
However, a recent measurement of isomeric yield ratios
pointed out that FF spin comes from the fission process and also CN angular momentum~\cite{Gjestvang,disentangle}.
Therefore, it is vital for theoretical studies to  disentangle the influences of angular momentum and
excitation energies on FF spin and scission by systematical variations of initial conditions.

The fission under rapid rotation is a very violent non-equilibrium non-adiabatic process~\cite{recent},
inducing strong spin alignment between fragments and CN.
In this respect,  the microscopic real-time dynamical approach is suitable.
TDDFT has been demonstrated to be a promising framework for descriptions of the later phase of fission after the saddle point~\cite{Goddard,Bulgac_real,Bulgac_saddle,SNature,superfluid,qy2}.
Even far from the equilibrium, the partition of energies, angular and linear momentum, and particles between nascent fragments can be self-consistently determined by TDDFT~\cite{Tanimura,Bulgac_shar,Unitary,qy4}.
Currently, within the TDDFT framework, the FF intrinsic spins are obtained by applying successive partial angular momentum projections (AMP).
However, only very low or zero angular momentum has been considered in earlier theoretical studies~\cite{BulgacL1,BulgacL2,ScampsC1}.
Besides, the high angular momentum is also relevant for heavy-ion reactions such as quasi-fission and multi-nucleon transfer reactions~\cite{ScampsC3,quasi}.
The objection of this Letter is to study the generation of FF spins and the scission mechanism across a wide range of initial angular
momentum and excitation energies
based on TDDFT, which has been a missing piece for a long time.

\emph{Methods.}---
The dynamical fission process after the saddle point of $^{240}$Pu is studied within the TDDFT framework,
specifically, the time-dependent Hartree-Fock+BCS (TD-BCS) approach~\cite{superfluid,formation}.
With a finite-temperature HF+BCS solution as an initial start for TD-BCS evolutions, the dynamical pairing effect
and the thermal excitation are self-consistently included~\cite{qy1,qy2}.
The initial wave functions are obtained by deformation constrained HF+BCS calculations at finite temperature~\cite{skyax, Goodman}.
Then the TD-BCS equation is solved as in Ref.~\cite{Ebata} using the modified Sky3D code~\cite{sky3d}.
In this work, the Skyrme force SkM* has been adopted~\cite{skm}, which has widely been used in  fission studies.
Although other forces like UNEDF1 and SLy4 are also suitable, they only result in minor changes in dynamical evolutions ~\cite{Bulgac_saddle}.
Besides, the mixing-type density dependent pairing interaction is adopted in our approach~\cite{MIX}.

Next the angular momentum of CN is obtained by an initial rotation of the system at the beginning of the evolution.
The system is rotated along the $y$-axis which is perpendicular to the splitting direction.
The driven rotation of the single-particle Hamiltonian is applied continuously for 30 fm/c at the beginning.
The Hamiltonian is written as,
\begin{equation}
\begin{aligned}
\hat{h}_{q}=&U_{q}(\vec{r})-\nabla\cdot[B_{q}(\vec{r})\nabla]+i\vec{W}_{q}\cdot(\vec{\sigma}\times\nabla)+\vec{S}_{q}\times\vec{\sigma} \\
&-\frac{i}{2}[(\nabla\times\vec{A}_{q})+2\vec{A}_{q}\cdot\nabla]
\end{aligned}
\end{equation}
where the mean-field potential $U_{q}(\vec{r})$, effective mass $B_{q}(\vec{r})$, spin-orbit potential $\vec{W}_{q}(\vec{r})$ and the time-odd terms $\vec{S}_{q}, \vec{A}_{q}$ are all set to rotate in a constant speed,
and $q$ denotes the isospin of neutron and proton.
For each time step $t_{i}$, the coordinate $\vec{r}$ of $[U_{q},B_{q},\vec{W}_{q},\vec{S}_{q},\vec{A}_{q}]$ rotates as
\begin{equation}
\vec{r}(t_{i})=(x_{i},y_{i},z_{i})\rightarrow (x_{i}+z_{i}cos(\omega\Delta t),y_{i},z_{i}-x_{i}sin(\omega\Delta t))
\end{equation}
The angular momentum can be adjusted by the rotating speed $\omega$.
Such an initial rotation method is similar to the cranking Hartree-Fock method,
which has been applied previously in TDHF studies~\cite{LuGuo,torus}.
However,
the cranking TDHF calculation is difficult to incorporate pairing effects and
particulary the initial thermal excitations.
Both the total energy and angular momentum of the CN system increase rapidly by the initial boost and then become stable.
To our knowledge, this work presents the first microscopic self-consistent study of CN fission dynamics
at rapid rotations and high excitations.

\emph{Results.}---
The fission evolutions of $^{240}$Pu based on TD-BCS with varying initial excitation energies and angular momentum
have been studied in a controllable way.
In our approach, the energy boosted by the initial rotation is almost collective rotation energy.
To study CN fission with energy dependencies,  initial configurations with temperatures at $T$=0.0, 1.0, 1.2 MeV
are adopted for TD-BCS evolutions.

\begin{figure}
	\centering
	\includegraphics[width=0.5\textwidth]{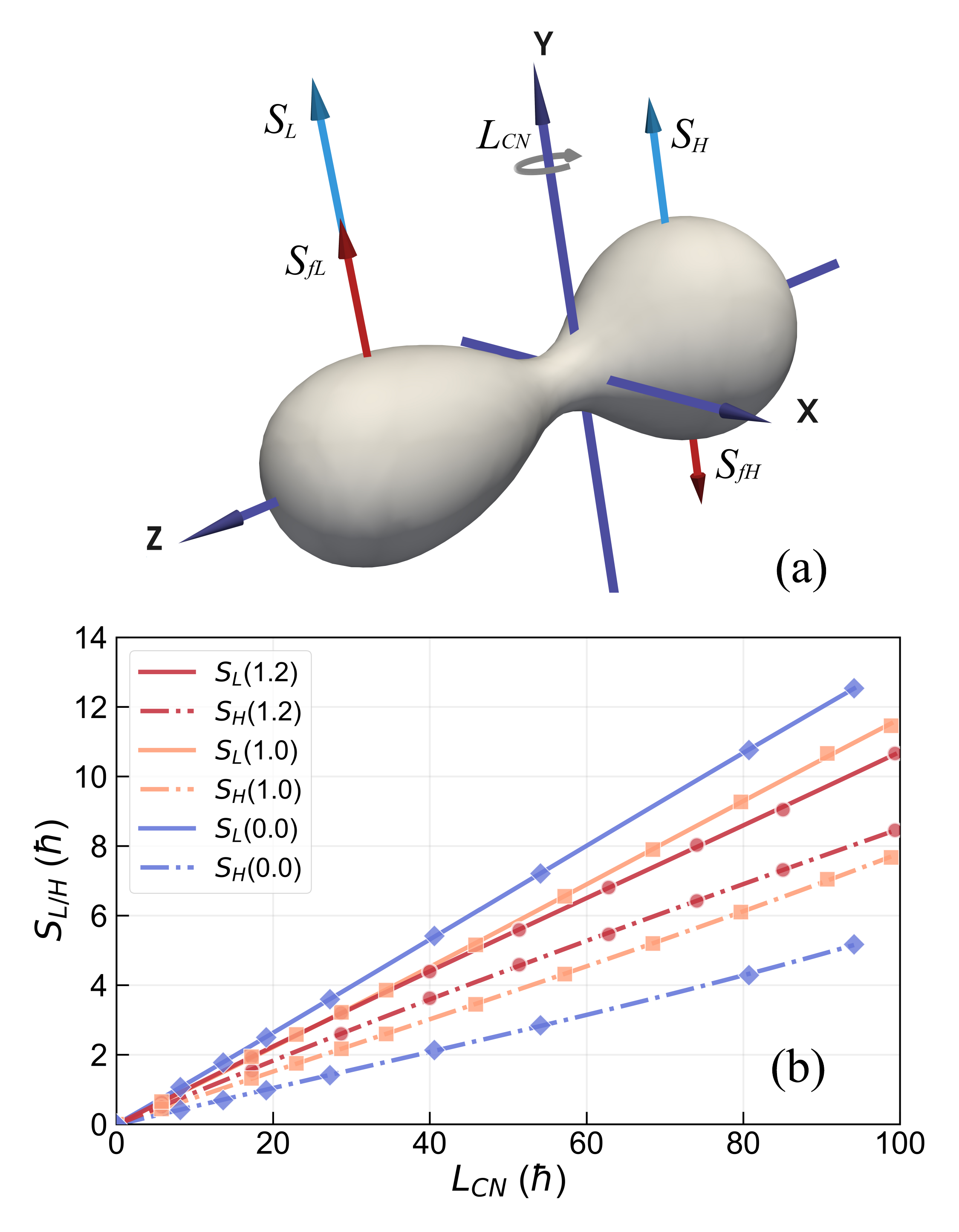}	
	\caption{ (a) Illustration of FF spin $S_{L/H}$ of a rotating CN   with angular momentum $L_{CN}$ along the $y$-axis, in which
fluctuation FF spins are denoted as $S_{fL}$ and $S_{fH}$ for light and heavy fragments, respectively.
   (b) Calculated spins of light and heavy FF as a function of $L_{CN}$ of $^{240}$Pu, with initial temperatures $T$=0.0, 1.0 and 1.2 MeV. }
	\label{Fig1}
\end{figure}

The orientations  of FF spins and the CN rotation are illustrated in Fig.\ref{Fig1}(a).
The FF spin along the $y$-axis can be calculated by $<J_y>$,  when two fragments are well separated~\cite{Bulgac_future}.
The obtained $<J_x>$ and $<J_z>$ are basically zero in all cases.
Note that AMP should be applied to obtain FF spin as CN at low angular momentum.
Under rapid rotations, the obtained average FF angular momentum is close to its intrinsic spin with AMP~\cite{crank_amp1,crank_amp2}.
As demonstrated in studies of quasi-fission fragment spins~\cite{ScampsC3}, the expectation value of the angular momentum
perpendicular to the reaction plane encapsulates most of the system's spin at high spins.
Furthermore, this expectation value increases toward the peak of the angular momentum distribution, also confirmed by cranking HF plus AMP calculations~\cite{crank_amp1,crank_amp2}.
The application of AMP at thermal excitations is too complicated.
Thus the present study focuses on the fission of CN with initial high angular momentum without AMP.

The calculated FF spin at different initial excitation energies and angular momentum are shown in Fig.\ref{Fig1}(b). It can be seen that both heavy and light fragment's average spins increase almost linearly with the increasing angular momentum of CN.
The spin of light fragment is larger than that of the heavy fragment, being consistent with the sawtooth structures~\cite{Wilson}.
The spin ratio of light to heavy fragments is unchanged up to high spins
but reduced as the temperature increases.
Finally the spins of two fragments become close at high temperatures. This implies
the fade away of sawtooth structures of FF spin at high excitations rather than at high angular momentum of CN.

It is known that the FF spin comprises a component $S_{r}$ inherited from the rotating CN and a component generated by fluctuations $S_{f}$ associated with six scission modes~\cite{RandrupC4,mode}.
The angular momentum of CN is shared by two fragments as $S_{rH/L}$ according to the momentum of inertia,
\begin{equation}
\frac{L_{CN}}{\mathcal{I}_{CN}}=\frac{S_{rH}}{\mathcal{I}_{H}}=\frac{S_{rL}}{\mathcal{I}_{L}}
\end{equation}
where $\mathcal{I}_{\text{CN}}$, $\mathcal{I}_{H}$, and $\mathcal{I}_{L}$ denote the moments of inertia of CN, the heavy and light fragments, respectively.
The rotation inertia can be calculated with the rigid inertia assumption, since nuclear superfluidity disappears quickly as CN angular momentum and excitation energies increase~\cite{collapse1,Goodman}.
Note that FF spin inherits only a very small portion of the CN angular momentum, since each FF inertia is much smaller than the CN inertia.
Generally the heavy fragment has a larger rigid inertia and  a higher inherited spin than the light fragment. This is different from the final partition of FF spin in Fig.\ref{Fig1}.
In contrast to spontaneous fission, where the intrinsic FF spin is governed by fluctuations, the scenario at high angular momentum is characterized by
 a combined contribution from inherited spin and from fluctuations.
To investigate the underlying scission modes, the component of FF spin generated by fluctuations must be isolated.

\begin{table}
\caption{
 Calculated fluctuation spins of heavy $S_{fH}$ and light $S_{fL}$ fragments of $^{240}$Pu, along with their portion ($P_{fH}$, $P_{fL}$) to the total FF spin.
Results for CN at different temperatures $T$  (MeV) and angular momentum $L_{CN}$ ($\hbar$) are displayed.
$E^{*}$ (MeV) denotes associated total excitation energies including rotational energies.
}
\begin{tabular}{l c c c c c}
 \hline
 \hline
 ~~$E^{*}(T)$  ~~&~~ $L_{CN}$ ~~&~~ $S_{fH}$ ~~&~~ $S_{fL}$ ~~&~~ $P_{fH}$ ~~&~~ $P_{fL}$~~\vspace{1pt}\\
 \hline
 1.8(0.0)  & 27.6   &  -1.04  &  1.47  & 70.1\%  &  41.8\%    \vspace{1pt} \\
 10.9(0.0) & 68.6   &  -2.67  &  3.42  & 78.3\%  &  40.0\%     \vspace{1pt}\\
 21.0(0.0) & 95.2   &  -3.41  &  4.64  & 70.3\%  &  40.2\%    \vspace{1pt} \\
 \hline
 20.9(1.0) & 28.8   &  -0.46  &  1.03  & 20.6\%  &  34.4\%    \vspace{1pt} \\
 30.7(1.0) & 68.7   &  -1.14  &  2.73  & 22.0\%  &  37.0\%    \vspace{1pt} \\
 41.1(1.0) & 99.3   &  -1.44  &  4.07  & 19.2\%  &  38.2\%   \vspace{1pt}  \\
 \hline
 34.4(1.2) & 28.8   &  -0.03  &  0.91  & 1.1\%   &  32.3\%    \vspace{1pt} \\
 40.6(1.2) & 63.0   &  -0.51  &  1.83  & 9.4\%   &  30.7\%    \vspace{1pt} \\
 52.6(1.2) & 99.3   &  -0.80  &  3.24  & 9.6\%   &  33.5\%    \vspace{1pt} \\
 \hline
 \hline
\end{tabular}
\label{Table1}
\end{table}

\begin{figure}
	\centering
	\includegraphics[width=0.5\textwidth]{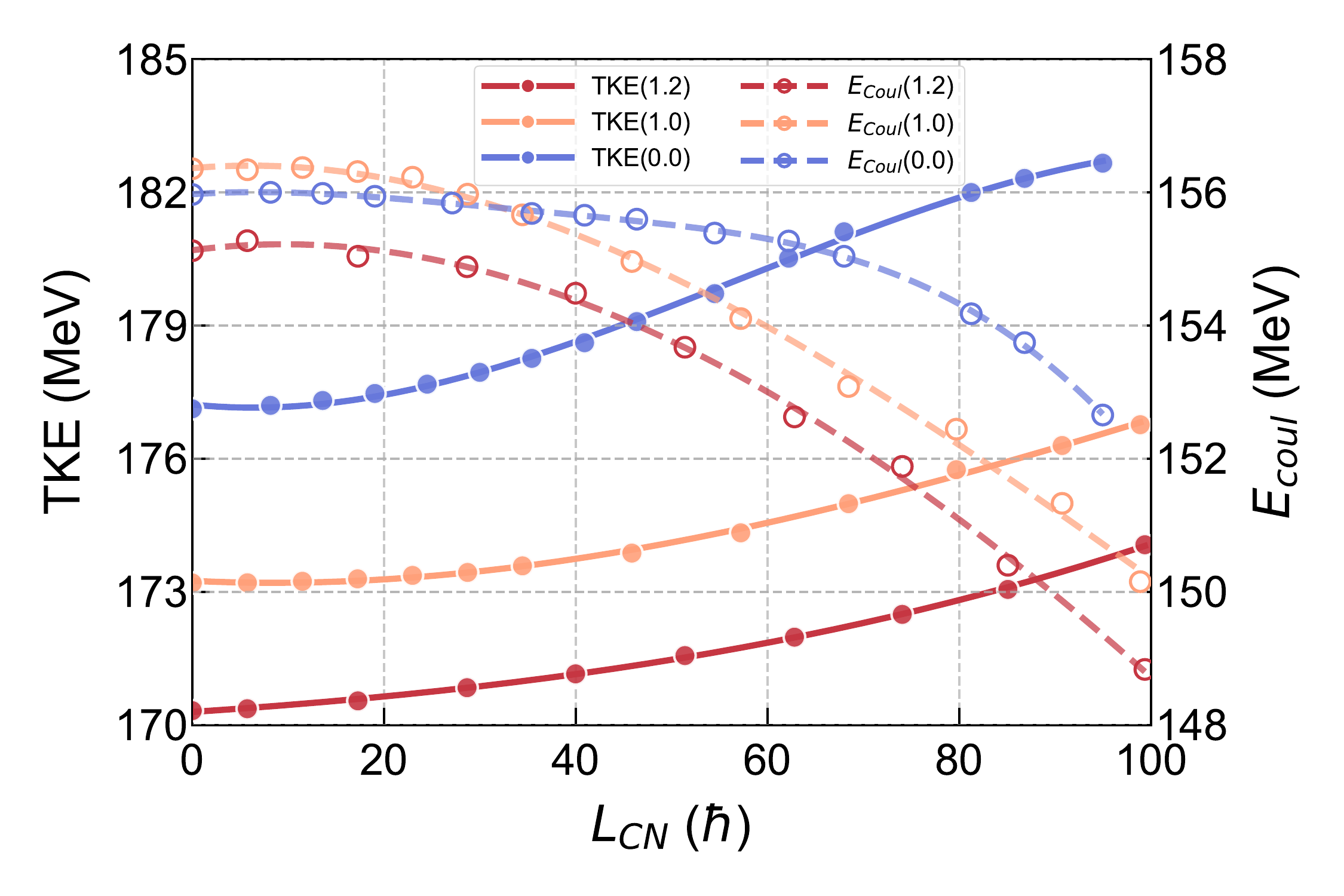}	
	\caption{Total kinetic energy (TKE) and extracted Coulomb energy $E_{Coul}$ between fission fragments as a function of $L_{CN}$ at initial temperatures of 0.0, 1.0, 1.2 MeV.
The Coulomb energy is determined when the neck reaches the same density for comparison.
 The solid lines denote TKE (the left vertical coordinate), and the dash-dotted lines denote the Coulomb energy (the right vertical coordinate). All energies are given in MeV.}
	\label{Fig2}
\end{figure}

The fluctuation spin $S_f$ can be obtained by subtracting the inherited FF spin $S_r$ from the total FF spin $S_{L/H}$,
as illustrated in Fig\ref{Fig1}.
Despite the general underestimation of fluctuation effects in TDDFT~\cite{stochastic}, this extracted net fluctuation component can avoid a discernible bias.
The obtained FF fluctuation spins are displayed in Table \ref{Table1} with varying CN angular momentum and $E^*$.
It can be seen that the fluctuation spins of light and heavy fragments are antiparallel along the $y$-axis.
Specifically, the $S_{fL}$ of light fragment is aligned with $L_{CN}$, while $S_{fH}$  is opposite to it.
This antiparallel FF fluctuation spins demonstrate clearly that the bending scission mode is dominated when CN has low initial temperatures~\cite{Equilibrium}.
For both heavy and light fragments, the fluctuation spin components $S_{f}$ increases with increasing $L_{\text{CN}}$.
Table \ref{Table1} also shows that the proportion of fluctuation spins in the total spin.
It can be seen that $S_{fH}$ decreases with increasing excitation energies.
The proportion of fluctuation spin of the heavy fragment $P_{fH}$ is very small at high excitations.
This indicates that the occurrence probability of the wriggling and bending scission modes are becoming equally important as the initial excitation energy increases.
The $S_{fL}$ is preferentially parallel to $L_{\text{CN}}$, whereas $S_{fH}$ is more likely to be anti-parallel to $L_{\text{CN}}$,
providing a counterbalance in the partition of spin and energy since the heavy fragment inherits more angular momentum.

The scission configuration can be significantly altered by the rapid rotation of CN, impacting multiple post-scission observables.
In Fig \ref{Fig2}, it is shown that TKE of fragments increase with the increasing $L_{CN}$, explaining the trend observed in the heavy-ion induced fission~\cite{S32}.
We also see that TKE decreases significantly with increasing initial $E^*$, owing to enhanced dissipation and elongated neck at high excitations~\cite{qy2}.
As a probe of the distance between fragments at scission, the Coulomb energies pre-scission are also shown in Fig \ref{Fig2}.
However, it shows that the Coulomb energies decreases significantly as $L_{CN}$ increases.
This indicates the rapid CN rotation leads to elongated neck configurations.
In non-rotating fission, the Coulomb energy is the major part of TKE.
Therefore, the increasing TKE as a function of $L_{CN}$ is attributed to relative rotational energies between fragments
instead of the Coulomb energy.
Or say that a large portion of the collective rotational motion of CN is converted into the relative motion between fragments or TKE after the scission.

\begin{figure}
	\centering
	\includegraphics[width=0.5\textwidth]{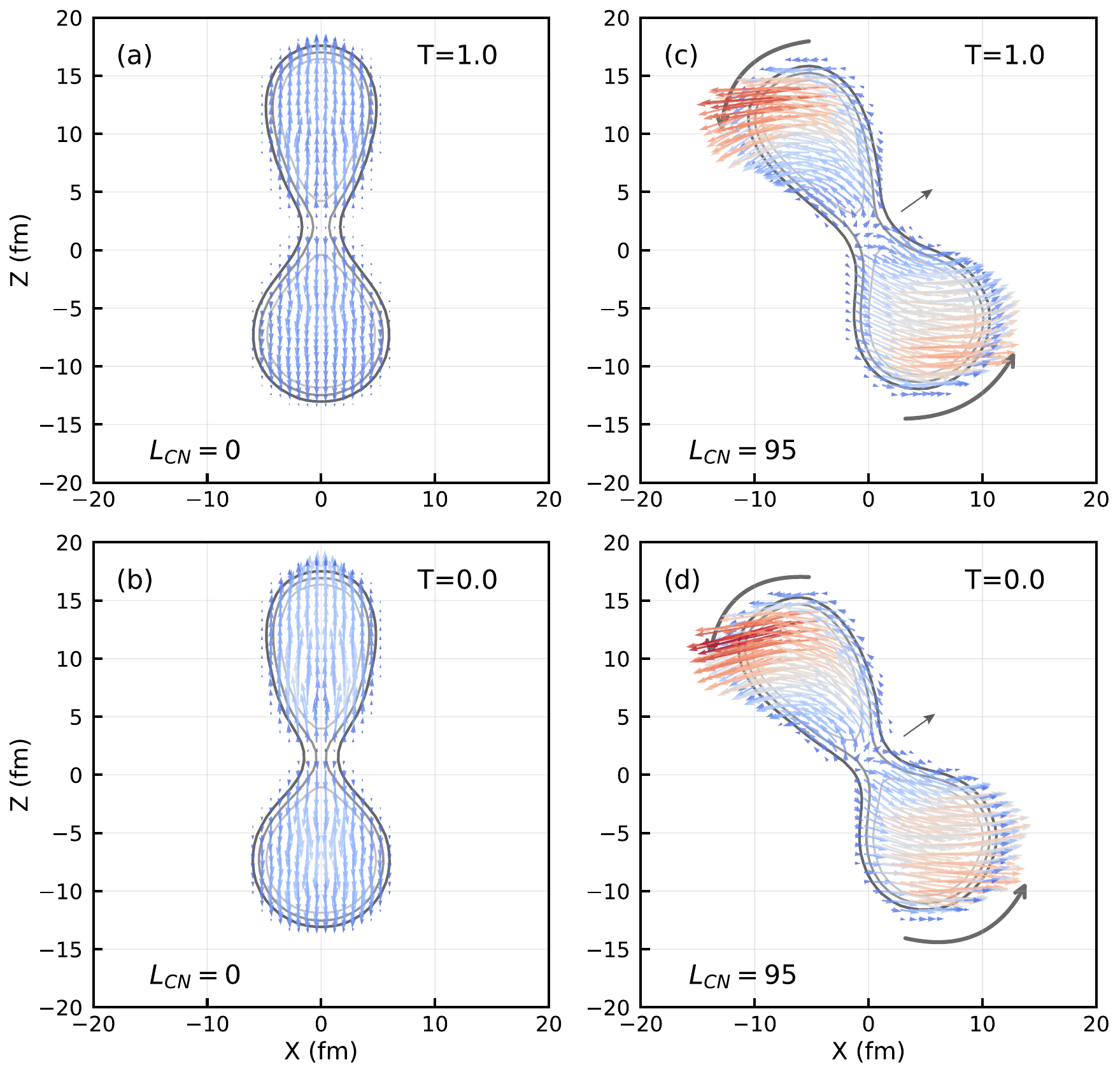}	
	\caption{
Calculated current density distributions ${\vec{j}}(\textbf{r})$ at scission with initial temperatures 0.0 and 1.0 MeV,  and $L_{CN}$=0 and 95$\hbar$, respectivley.
Neck currents show scission nucleons emitted perpendicular to the fission axis.
Density contours are shown with quadrupole deformation $\beta_{2}=4.0$ for comparison. }
	\label{Fig3}
\end{figure}

To illustrate the scission configurations with varying initial angular momentum and excitation energies,  Fig. \ref{Fig3} shows the evolution of pre-scission
current density ${\vec{j}}(\textbf{r})$.
The currents, related to velocity fields, exhibit violent fission dynamics at rapid rotation.
Without initial rotations, the scission neck becomes thicker at higher excitation energies due to
 enhanced viscosity~\cite{viscosity}.
As shown in density contours, it can be seen that neck sizes become thicker under rapid rotations irrespective of initial temperatures.
The neck is particulary thicker under highly-excited rapid rotations.
These neck sizes are determined at the same quadrupole deformation $\beta_{2}=4.0$ for comparison.
The thicker neck size allows an elongated scission configuration and consequently
the decreased Coulomb energy, being consistent with Fig.\ref{Fig2}.
This can be understood that the strong centrifugal effect leads to
a violent non-adiabatic splitting that stretches the neck.
Experimentally, the widths of fragment mass distributions increase at high excitation energies and high angular momentum~\cite{S32}, demonstrating the influences of thicker and elongated neck configurations.
Furthermore, we see clearly that there are scission currents with a polarized direction being perpendicular to the fission axis at high spins.
This indicates that scission nucleons emitted, mainly neutrons, are in the direction perpendicular to the fission axis at high $L_{CN}$,
which is not presented at zero $L_{CN}$.
An earlier study shows that at low excitations, scission neutrons emitted in roughly equal numbers in the equatorial plane and along the fission axis~\cite{Bulgac_neck}.
Thus scission neutrons in rapidly rotating fission are distinctly different from low energy fission, which can be verified by heavy-ion induced fission.
Besides, as evidenced by density contours at $L_{CN}=95 \hbar$, the light fragment displays pronounced triaxial deformation~\cite{torque}.
Note that the fragment shapes are related to the distortion energies, and also the sharing of excitation energies and intrinsic spins~\cite{HanR,ScampsGB}.

\emph{Conclusion.}---
We have self-consistently incorporated initial rotation and thermal excitations
 into  microscopic real-time descriptions of nuclear fission process, offering insights into the scission mechanism
 and the generation of FF spin.
For fission of $^{240}$Pu under rapid rotation, the light FF spin is larger than
 the heavy FF spin, and the spin ratio is unchanged with increasing CN angular momentum.
 However, the partition of FF spin is dependent on excitation energies and
 the spins of two fragments become close at high excitation energies, indicating the
 disappearance of sawtooth structures in FF spins at high excitations rather than high angular momentum.
Actually the heavy fragment inherits more angular moment with a larger inertia.
 Further analysis finds fluctuation spins of fragments are antiparallel at low excitations,
 exhibiting the bending scission mode. The mixture of different scission modes  are likely to happen at high excitation
 energies.

 The obtained TKE and the inherited angular momentum also shows that
 most of the CN angular momentum is converted into relative motion between fragments.
 The Coulomb energies before scission decrease with increasing angular momentum.
 This is consistent with enhanced neck sizes and elongated scission deformations under rapid rotations, explaining experimental
 observations of increased TKE and widths of fragment mass distributions in heavy-ion induced fission.
 In addition, due to rapid rotations, there are scission neutron emissions in the direction perpendicular to the fission axis,
 which is different from the scenario in low energy fission.
This work demonstrates that the violent non-equilibrium non-adiabatic scission
under rapid rotations can significantly impact post-fission observables.
In conclusion, the partition ratio of FF spins is sensitive to excitation energies and not dependent on
initial angular momentum, however, the scission mechanism and  post-fission
 observables are dependent on both conditions.

\section*{Acknowledgements}
We thank useful discussions with F.R. Xu and H.M. Jia.
 This work was supported by  the
 National Key R$\&$D Program of China (Grant 2023YFE0101500, No.2023YFA1606403),
  the National Natural Science Foundation of China under Grants No.12475118, 12335007.





\begin{thebibliography}{99}

\bibitem{exp_frag_spin1}
J. B. Wilhelmy, E. Cheifetz, R. C. Jared, S. G. Thompson, H. R. Bowman, and J. O. Rasmussen,
Angular Momentum of Primary Products Formed in the Spontaneous Fission of $^{252}\mathrm{Cf}$,
Phys. Rev. C 5, 2041 (1972).

\bibitem{clouds}
P.Colciaghi, Y. Li, P. Treutlein, and T. Zibold,
Einstein-Podolsky-Rosen Experiment with Two Bose-Einstein Condensates,
Phys. Rev. X 13, 021031(2023).

\bibitem{Wilson}
J. Wilson et al.,
Angular momentum generation in nuclear fission,
Nature  590, 566 (2021).

\bibitem{NeutSaw1}
J. Terrell,
Neutron yields from individual fission fragments,
Phys. Rev. 128, 2925 (1962).

\bibitem{NeutSaw2}
C. Budtz-Jorgensen and H.-H. Knitter,
Simultaneous investigation of fission fragments and neutrons in $^{252}\mathrm{Cf}$(SF),
Nucl. Phys. A 490, 307 (1988).

\bibitem{carlsson}
M. Albertsson, B. G. Carlsson, T. D{\o}ssing, P. M\"{o}ller, J. Randrup, and S. {\AA}berg,
Excitation energy partition in fission,
Phys. Lett. B 803, 135276 (2020).

\bibitem{StetcuL}
I. Stetcu, A. E. Lovell, P. Talou, T. Kawano, S. Marin, S. A. Pozzi, and A. Bulgac,
Angular Momentum Removal by Neutron and $\gamma$-Ray Emissions During Fission Fragment Decays,
Phys. Rev. Lett. 127, 222502 (2021).

\bibitem{RandrupL}
J. Randrup and R. Vogt,
Generation of Fragment Angular Momentum in Fission,
Phys. Rev. Lett. 127, 062502 (2021).

\bibitem{RandrupC1}
J. Randrup,
Coupled fission fragment angular momenta,
Phys. Rev. C 106, L051610 (2022).

\bibitem{RandrupC4}
J. Randrup, and R. Vogt,
Refined treatment of angular momentum in the event-by-event fission model FREYA,
Phys. Rev. C 89, 044601 (2014).

\bibitem{BulgacL1}
A. Bulgac, I. Abdurrahman, S. Jin, K. Godbey, N. Schunck, and I. Stetcu,
Fission Fragment Intrinsic Spins and Their Correlations,
Phys. Rev. Lett. 126, 142502 (2021).

\bibitem{BulgacL2}
A. Bulgac, I. Abdurrahman, K. Godbey, and I. Stetcu,
Fragment Intrinsic Spins and Fragments' Relative Orbital Angular Momentum in Nuclear Fission,
Phys. Rev. Lett. 128, 022501 (2022).

\bibitem{BulgacC1}
A. Bulgac,
Angular correlation between the fission fragment intrinsic spins,
Phys. Rev. C 106, 014624 (2022).

\bibitem{ScampsC1}
G. Scamps, I. Abdurrahman, M. Kafker, A. Bulgac, I. Stetcu et al.
Spatial orientation of the fission fragment intrinsic spins and their correlations,
Phys. Rev. C 108, L061602 (2023).

\bibitem{ScampsC2}
G. Scamps,
Quantal effect on the opening angle distribution between the spins of the fission fragments,
Phys. Rev. C 109, L011602 (2024).

\bibitem{photon_exp}
J. Randrup, D. Thomas, and R. Vogt,
Probing fission fragment angular momenta by photon measurements,
Phys. Rev. C 106, 014609 (2022).

\bibitem{brosa}
U. Brosa, S. Grossmann, and A. M{u}ller,
Nuclear scission,
Physics Reports 197, 167 (1990)

\bibitem{scission_point}
B. D. Wilkins, E. P. Steinberg, and R. R. Chasman,
Scission-point model of nuclear fission based on deformed-shell effects,
Phys. Rev. C 14, 1832 (1976)

\bibitem{qy4}
Y. Qiang, J. C. Pei, and K. Godbey,
Quantum entanglement in nuclear fission,
Phys. Lett. B 861, 139248 (2025).

\bibitem{shang}
H.Y. Shang, Y. Qiang, J.C. Pei,
Energy partition between entangled fission fragments,
NUCL SCI TECH 36, 211 (2025).

\bibitem{mode}
R.P. Schmitt, A. J. Pacheco,
Equilibrium treatment of spin-depolarizing modes in mass asymmetric heavy-ion systems,
Nucl. Phys. A 379, 313-329 (1982).

\bibitem{mode2}
T. D\o ssing, J. Randrup,
Dynamical evolution of angular momentum in damped nuclear reactions: (I). Accumulation of angular momentum by nucleon transfer,
Nucl. Phys. A 433, 215 (1985).

\bibitem{freya_mode}
R. Vogt, and J. Randrup,
Observational Consequences of Angular Momentum in Fission,
EPJ Web of Conferences 322, 07001 (2025).

\bibitem{Marevic1}
P. Marevic, N. Schunck, J. Randrup, and R. Vogt,
Angular momentum of fission fragments from microscopic theory,
Phys. Rev. C 104, L021601 (2021).

\bibitem{RandrupC2}
T. D\o{}ssing, S. \AA{}berg, M. Albertsson, B. G. Carlsson, and J. Randrup,
Angular momentum in fission fragments,
Phys. Rev. C 109, 034615 (2024).

\bibitem{qy2}
Y. Qiang and J. C. Pei,
Energy and pairing dependence of dissipation in real-time fission dynamics,
Phys. Rev. C 104, 054604(2021).


\bibitem{qy1}
Y. Qiang, J. C. Pei, and P. D. Stevenson,
Fission dynamics of compound nuclei: Pairing versus fluctuations,
Phys. Rev. C 103, L031304(2021).



\bibitem{Gjestvang}
D. Gjestvang et al.,
Examination of how properties of a fissioning system impact isomeric yield ratios of the fragments,
Phys. Rev. C 108, 064602 (2023).

\bibitem{disentangle}
C. Simone et al.,
Disentangling the influence of excitation energy and compound nucleus angular momentum on fission fragment angular momentum,
Phys. Rev. C 111, L031601 (2025).

\bibitem{recent}
R. Choudhury,
Recent studies in heavy ion induced fission reactions,
Pramana-J. Phys. 57, 585 (2001).

\bibitem{Goddard}
P. Goddard, P. Stevenson, and A. Rios,
Fission dynamics within time-dependent Hartree-Fock: Deformation-induced fission,
Phys. Rev. C 92, 054610 (2015).

\bibitem{Bulgac_real}
A. Bulgac, P. Magierski, K. J. Roche, and I. Stetcu,
Induced Fission of 240Pu within a Real-Time Microscopic Framework,
Phys. Rev. Lett. 116, 122504 (2016).

\bibitem{Bulgac_saddle}
A. Bulgac, S. Jin, K. J. Roche, N. Schunck, and I. Stetcu,
Fission dynamics of 240Pu from saddle to scission and beyond,
Phys. Rev. C 100, 034615 (2019).

\bibitem{SNature}
G. Scamps and C. Simenel,
Impact of pear-shaped fission fragments on mass-asymmetric fission in actinides,
Nature 564, 382 (2018).

\bibitem{superfluid}
G. Scamps, C. Simenel, and D. Lacroix,
Superfluid dynamics of 258Fm fission,
Phys. Rev. C 92, 011602(R) (2015).

\bibitem{Tanimura}
Y. Tanimura, D. Lacroix, S. Ayik,
Microscopic Phase-Space Exploration Modeling of 258Fm Spontaneous Fission,
Phys. Rev. Lett. 118, 152501(2017).

\bibitem{Bulgac_shar}
A. Bulgac,
Fission-fragment excitation energy sharing beyond scission,
Phys. Rev. C 102, 044609 (2020).

\bibitem{Unitary}
A. Bulgac, S. Jin, and I. Stetcu,
Unitary evolution with fluctuations and dissipation,
Phys. Rev. C 100, 014615 (2019).

\bibitem{ScampsC3}
G. Scamps,
Microscopic study of spin transfer in near-barrier nuclear reactions,
Phys. Rev. C 110, 054605 (2024).

\bibitem{quasi}
H. Lee, P. McGlynn, and C. Simenel
Shell effects in quasifission in reactions forming the $^{226}\mathrm{Th}$ compound nucleus,
Phys. Rev. C 110, 024606 (2024).

\bibitem{formation}
C. Simenel and A. S. Umar,
Formation and dynamics of fission fragments,
Phys. Rev. C 89, 031601 (2014).

\bibitem{skyax}
P.-G. Reinhard, B. Schuetrumpf, and J. A. Maruhn,
The Axial Hartree-Fock + BCS Code SkyAx,
Comp. Phys. Comm. 258, 107603 (2021).

\bibitem{Goodman}
A. L. Goodman, Nucl. Phys. A 352 (1981) 30.

\bibitem{Ebata}
S. Ebata, T. Nakatsukasa, T. Inakura, K. Yoshida, Y. Hashimoto, and K. Yabana,
Canonical-basis time-dependent Hartree-Fock Bogoliubov theory and linear-response calculations,
Phys. Rev. C 82, 034306 (2010).

\bibitem{sky3d}
J. A. Maruhn, P.-G. Reinhard, P. D. Stevenson, and A. S. Umar,
The TDHF code Sky3D,
Comp. Phys. Comm. 185, 2195(2014).

\bibitem{skm}
J. Bartel, P. Quentin, M. Brack, C. Guet, and H.-B. H{\aa}kansson,
Towards a better parametrisation of Skyrme-like effective forces: A critical study of the SkM force,
Nuclear Physics A 386, 79 (1982).

\bibitem{MIX}
J. Dobaczewski, W. Nazarewicz, and M. V. Stoitsov,
Nuclear ground-state properties from mean-field calculations,
Eur. Phys. J. A 15, 21 (2002).

\bibitem{LuGuo}
L. Guo, J. A. Maruhn, P.-G. Reinhard, and Y. Hashimoto,
Conservation properties in the time-dependent Hartree Fock theory,
Phys. Rev. C 77, 041301 (2008).

\bibitem{torus}
T. Ichikawa, K. Matsuyanagi, J. A. Maruhn, and N. Itagaki,
High-spin torus isomers and their precession motions,
Phys. Rev. C 90, 034314 (2014).

\bibitem{Bulgac_future}
A. Bulgac, S. Jin, and I. Stetcu,
Nuclear fission dynamics: Past, present, needs, and future,
Front. Phys. 8, 63 (2020).

\bibitem{crank_amp1}
D. Baye, and P. H. Heenen,
Angular momentum projection on a mesh of cranked Hartree-Fock wave functions,
Phys. Rev. C 39,1056(1984).

\bibitem{crank_amp2}
H. Zdu\'{n}czuk, W. Satu{\l}a, J. Dobaczewski, and M. Kosmulski,
Angular momentum projection of cranked Hartree-Fock states: Application to terminating bands in $A\sim44$ nuclei,
Phys. Rev. C 76, 044304 (2007).


\bibitem{collapse1}
W. Nazarewicz, R. Wyss, and A. Johnson,
Structure of superdeformed bands in the A$\approx$150 mass region,
Nucl. Phys. A 503, 285(1989).

\bibitem{stochastic}
D. Lacroix, and S. Ayik,
Stochastic quantum dynamics beyond mean field,
Eur. Phys. J. A 50, 95 (2014).

\bibitem{Equilibrium}
R. P. Schmitt and A. J. Pacheco,
Equilibrium treatment of spin-depolarizing modes in mass asymmetric heavy-ion systems,
Nuclear Physics A 379, 313 (1982).

\bibitem{S32}
B. G. Glagola, B. B. Back, and R. R. Betts,
Effects of large angular momenta on the fission properties of Pt isotopes,
Phys. Rev. C 29, 486 (1984).

\bibitem{viscosity}
K. T. R. Davies, A. J. Sierk, and J. R. Nix,
Effect of viscosity on the dynamics of fission,
Phys. Rev. C 13, 2385 (1976).


\bibitem{Bulgac_neck}
I. Abdurrahman, M. Kafker, A. Bulgac, and I. Stetcu,
Neck Rupture and Scission Neutrons in Nuclear Fission,
Phys. Rev. Lett. 132, 242501 (2024).

\bibitem{torque}
G. Scamps,
Microscopic description of the torque acting on fission fragments,
Phys. Rev. C 106, 054614 (2022).

\bibitem{HanR}
R. Han, M. Warda, A. Zdeb, and L. M. Robledo,
Scission configuration in self-consistent calculations with neck constraints,
Phys. Rev. C 104, 064602 (2021).

\bibitem{ScampsGB}
G. Scamps and G. Bertsch,
Generation, dynamics, and correlations of the fission fragments' angular momenta,
Phys. Rev. C 108, 034616 (2023).















\end{thebibliography}

\end{document}